R. Kusche*, P. Klimach, A. Malhotra, S. Kaufmann, and M. Ryschka*

# An in-ear pulse wave velocity measurement system using heart sounds as time reference

**Abstract:** Pulse wave measurements provide vital information in medical diagnosis. For this reason, a measurement system is developed for determining the transient time of the pulse wave between the heart and the ear. To detect pressure variations in the sealed ear canal, caused by the arriving pulse wave, an in-ear sensor is developed which uses heart sounds as time reference. Furthermore, for extracting the heart sounds from the pressure measurements and calculating the pulse wave transient time, a MATLAB-based algorithm is described. An embedded microcontroller based measurement board is presented, which realizes an interface between the sensor and the computer for signal processing.

## 1 Introduction

Non-invasive and portable patient monitoring has become increasingly important. One interesting patient monitoring parameter is the stiffness of the arteries, especially of the aorta [1, 2]. In addition to invasive measurement methods, the stiffness can be determined via the Pulse Wave Velocity (PWV), which increases with the arterial stiffness. The PWV can be determined by analysis of the pulse wave morphology or by direct measurement of the travel time between two distant points, or between the heart and another peripheral point [1–5]. In a human body the PWV is usually in the range of 4 m/s and 9 m/s [3]. If the distance between the heart and the measurement point is known, it is sufficient to measure the Pulse Transient Time (PTT).

This work proposes a method to determine the PTT between the heart and the ear of a patient. The two-point measurement method which designates the heart sounds as the starting time of the Pulse Wave (PW) and the pressure changes in the sealed ear canal as arrival time, is executed directly at the patient's ear with a wearable embedded microcontroller-based system. It needs no further sensors at any other point of the human body. The developed measurement system also provides Electrocardiogram (ECG) and Photoplethysmography (PPG) circuits. The acquired data can be sent via a Bluetooth or a Universal Serial Bus (USB) connection to a host computer to display the measurement data. Additionally, the data can be stored on a microSD card.

**\*Corresponding Author: R. Kusche:** Laboratory of Medical Electronics, Lübeck University of Applied Sciences, Germany, E-mail: roman.kusche@fh-luebeck.de and Graduate School for Computing in Medicine and Life Sciences, Universität zu Lübeck, Germany
**\*Corresponding Author: M. Ryschka:** Laboratory of Medical Electronics, Lübeck University of Applied Sciences, Germany, E-mail: martin.ryschka@fh-luebeck.de
**P. Klimach:** Laboratory of Medical Electronics, Lübeck University of Applied Sciences, Germany and Graduate School for Computing in Medicine and Life Sciences, Universität zu Lübeck, Germany
**A. Malhotra:** Graduate School for Computing in Medicine and Life Sciences, Universität zu Lübeck, Germany and Institute of Medical Engineering, Universität zu Lübeck, Germany
**S. Kaufmann:** Laboratory of Medical Electronics, Lübeck University of Applied Sciences, Germany

## 2 Measurement methods

### 2.1 Electrocardiogram (ECG)

The electrocardiogram reflects the electric activities of the heart. It can be used as a time reference for heartbeat dependent physiological measurements. A simple ECG signal is easy to acquire and contains easy accessible timing in-formation.

### 2.2 Photoplethysmography (PPG)

Photoplethysmography is widely used for the implementation of pulse oximeters to determine the oxygen saturation of the blood [6]. In this work it is implemented as an additional time reference to compare the in-ear pulse wave with the changes in blood flow at a position close to the ear. Therefore, a reflective system with only one wavelength of approximately 940 nm is used which is actually sensitive to the blood volume within the illuminated region [7].

### 2.3 Heart sounds (HS)

Heart Sounds occur due to the mechanical activities during the heartbeat cycle. For this application the first heart sound (S1) is especially interesting because it can be used as a time reference for the opening of the aortic valve [7] which can be regarded as the starting point of the pulse wave. In comparison to the PWV the speed of the sound signals inside the body is so high that any propagation delay can be neglected.

### 2.4 In-ear pressure

The detection of the pulse wave inside the sealed ear canal has already been reported [8, 9]. The measurement principle is shown in Figure 1. Pulse wave induced movements in the ear canal seem to cause volume changes. These volume changes can be detected as pressure changes inside the sealed canal.

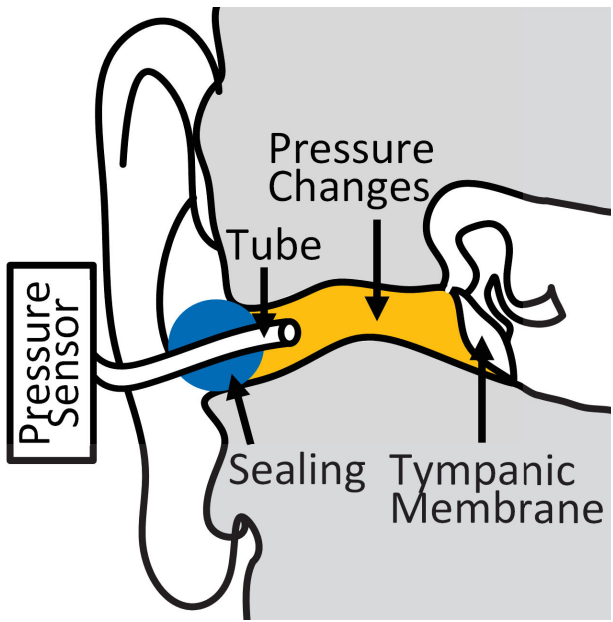

**Figure 1:** Measurement principle of detecting the pulse wave in the sealed ear canal.

### 2.5 Pulse transient time (PTT) and pulse wave velocity (PWV)

In a simple basic model the arterial elasticity ($E$) is correlated to the PWV by the Moens-Korteweg equation (1), where $h$ denotes the arterial wall thickness, $r$ its radius and $\rho$ the blood density [3, 10].

$$\text{PWV} \approx \sqrt{\frac{Eh}{2r\rho}} \quad (1)$$

In order to measure the PWV, it is sufficient to measure the PTT if the travel distance $\Delta x$ is known. As already presented [8, 9], the pressure pulse wave detected inside the ear canal can be used to detect the arrival time of the pulse wave. Formerly, the starting time of the pulse wave from the heart was derived using the ECG. However, the ECG only displays the electrical activity of the heart and not the aortic valve opening which is delayed by the Pre-Ejection-Period (PEP) with respect to R-peak of the ECG [11]. Since it is not constant, and shows some patient dependent variations, the PEP introduces some uncertainty to the PWV measurement. In order to reduce this uncertainty, the first heart sound (S1) is used as a time reference instead of the ECG signal. Figure 2 shows an example of the correlation of the Electrocardiogram (ECG), Heart Sounds (HS) and pressure Pulse Wave (PW) signals. It demonstrates that the usage of the heart sounds eliminates the influence of the pre-ejection-period. Therefore, there is no need of acquiring an ECG signal.

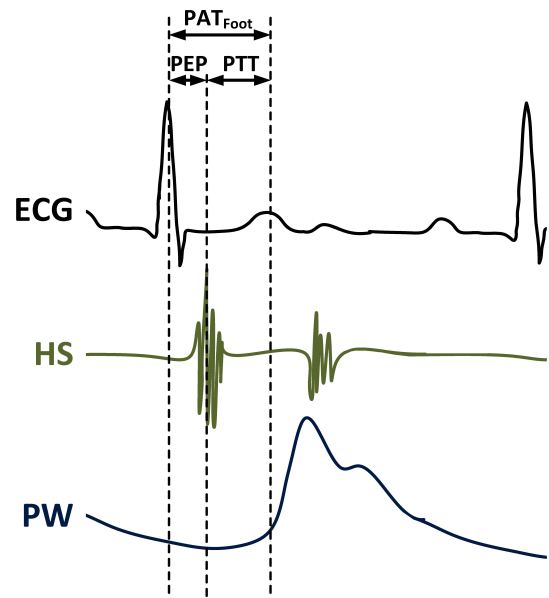

**Figure 2:** Exemplary timing relationships between the Electrocardiogram (ECG), Heart Sounds (HS) and Pulse Wave (PW).

The PWV can finally be calculated with formula (2) where $\Delta x$ denotes the distance between the heart and the ear [10].

$$PWV = \frac{\Delta x}{PAT - PEP} = \frac{\Delta x}{PTT} \quad (2)$$

# 3 Implementation

## 3.1 Sensor system

Since both the pulse wave inside the sealed ear canal, as well as the heart sounds are pressure changes in general, just one type of sensor is needed to acquire both signals. In this work an electret microphone capsule is used. To enable the measurement of the pulse wave's low frequency components in addition to the heart sounds, the microphone is modified to a differential pressure sensor. As depicted in Figure 3, an ear plug of a stethoscope is used to seal the ear canal against the ambient pressure and to mount the microphone capsule. By sealing the microphone membrane against the ear canal in principal a low frequency sensitive differential pressure sensor setup is achieved.

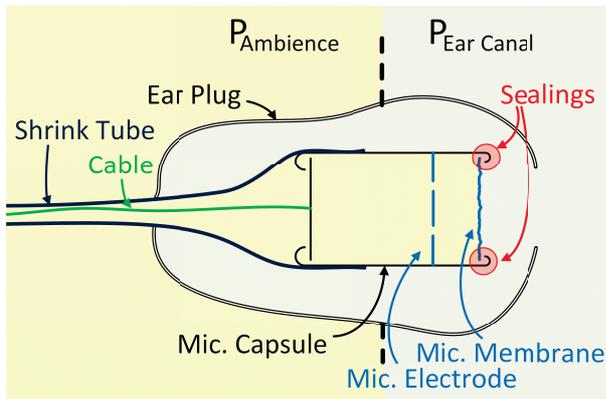

**Figure 3:** Functional block diagram of the developed sensor system to acquire the pulse wave and the heart sounds inside the ear.

## 3.2 Block diagram

Figure 4 shows the principle block diagram of the developed measurement system.

It consists of analogue circuits (ECG, Microphone Ampli-fiers, PPG), whose output signals are digitised with a 6 channel, 24 bit Analogue-to-Digital Converter (ADC, ADS131E06 from Texas Instruments) with a sample rate of 1000 Samples Per Second (SPS). The 32 bit microcontroller (ATSAM4S16C from Atmel) transmits the data via a USB or Bluetooth connection (LMX9838 from Texas Instruments) to the host computer. Additionally, the data can be stored on a microSD card. The System can be powered by a rechargeable lithium-ion battery or via a medical power supply which is in compliance with the IEC60601-1. For fur-ther applications an acceleration sensor is implemented which can e.g. be used to determine the patient's position and movements.

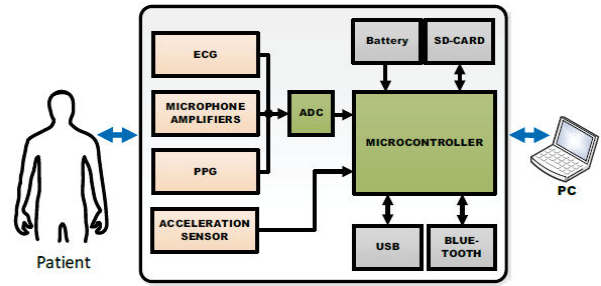

**Figure 4:** Principle block diagram of the developed measurement system to acquire the PW and the heat sounds inside the ear canal.

## 3.3 Electrocardiogram (ECG) module

The implemented ECG circuit uses an instrumentation amplifier (INA126E from Texas Instruments) as an input amplifier with a Driven Right Leg circuitry and shield driver. The signal is band pass limited with cut-off frequencies at 0.2 Hz and 1.7 kHz respectively.

## 3.4 Microphone amplifiers

To acquire data from up to four microphones, four separate microphone amplifiers are provided. Each of these four circuits consists of a highpass filter, an amplifier and a lowpass filter. To reduce the DC component of the microphone signal, a passive high pass of $1^{st}$ order is used. The lowpass after the amplification works as an anti-aliasing filter.

## 3.5 Photoplethysmogram (PPG) module

Since the PPG is an additional time reference, which is not needed in the normal PWV measurement process, it is realized in a very simple way. It consists of a transimpedance converter and an anti-aliasing filter. The transimpedance converter is implemented with an operational amplifier (OPA380 from Texas Instruments).

## 3.6 Microcontroller system and communication interfaces

As shown in Figure 4, the microcontroller is connected to all measurement components and controls the commu-nication interfaces. It is a 32 bit controller with an ARM Cortex-M4 core, clocked with a frequency of 120 MHz. In addition, it has a Digital Signal Processing (DSP) core. The controller's major task is the configuration of the ADC and the transmission of the digitized data to a PC via a USB connection or wireless via the Bluetooth module. Storing the data on a microSD card is executed via the controller's Serial Peripheral Interface (SPI).

### 3.7 Power supply

The measurement system can be powered with an external medical power supply. For wearable usage, a lithium-ion battery is preferable. An implemented battery controller is able to charge the battery and it provides under voltage protection. The system's power dissipation is about 670 mW, depending on the measurement setup. The battery's output voltage is 3.7 V and it has a maximum charge of approximately 1.2 Ah. This results in a battery lifetime of up to 6.5 hours.

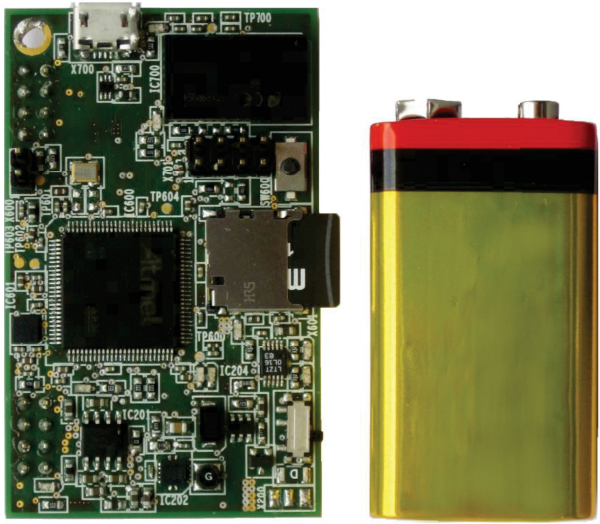

**Figure 5:** Manufactured and populated 6-layer PCB of the measurement system. The PCB contains more than 200 components. The 9 V block battery is shown for comparing the PCB's dimensions.

### 3.8 Manufactured system

The manufactured prototype is shown in Figure 5. The Printed Circuit Board (PCB) has dimensions of about $61 \times 37$ mm$^2$ and contains more than 200 components. To achieve this high packing density, the layout consists of 6 conductive layers and uses double side component population.

### 3.9 Algorithm

The current PC based signal algorithm is implemented using MATLAB. It extracts the information of the PTT from the raw data of the digitized microphone signal. In Figure 6 a simplified block diagram of the signal processing is shown. Since the amplitudes of the heart sounds are very small in comparison to the pulse wave's amplitudes, there is no filtering of the pulse wave signal necessary.

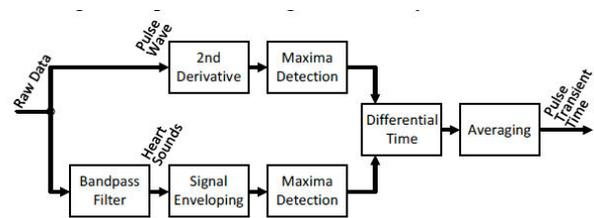

**Figure 6:** Block diagram of the algorithm to determine the PTT.

After calculation of the 2$^{nd}$ derivative of the pulse wave signal, the maxima are detected to determine the temporal beginning of the pulse wave. The heart sounds are extracted from the raw data with a very frequency-selective bandpass filter. Afterwards, the sound signal is enveloped by a Hilbert filter. The subsequent maxima detection block calculates the discrete point in time which represents the maximum amplitude of the heart sound. The resulting differential time of both signals is taken as the PTT. To improve the results of the algorithm, averaging over some heart beats is preferable.

## 4 Results and discussion

Figure 7 gives the first measurement results. The signals are taken from a young healthy male subject. The signal at the top (blue) is the acquired ECG signal followed by the raw data from a microphone amplifier (green, InEar), which is connected to the in-ear sensor, the second derivative of the in-ear signal (red, InEar 2nd Deri.), and the filtered heart sound signal (purple, InEar Filtered) at the bottom.

To highlight the derived PTTs, orange rectangles are drawn in the plot. In this case the pulse transient times are about 65 ms. Assuming a distance of $\Delta x = 0.35$ m between the heart and the ear of the subject with equation (2) the measured PTT results in a PWV of about 5.4 m/s, which in comparison with literature values seems to be quite realistic [3].

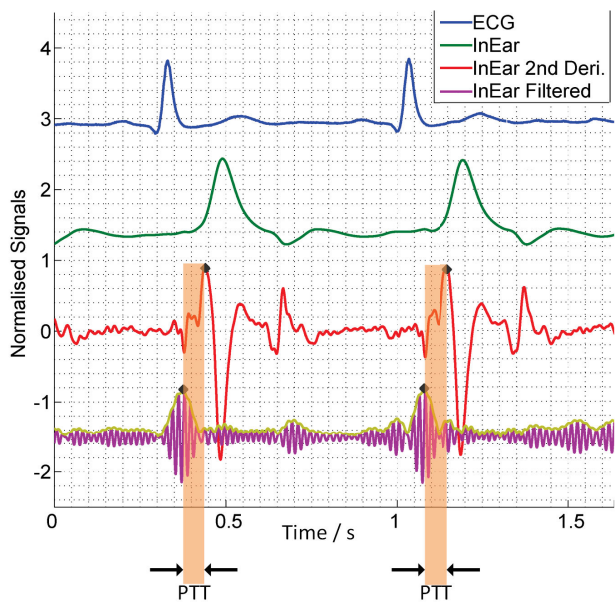

**Figure 7:** First measurement results. (Normalised signals).

# 5 Conclusion

A new measurement method to determine the PWV with just a single in-ear sensor was developed and its functionality is demonstrated with first measurements. In addition to the concept, a miniaturised embedded system has been built to test the proposed method in everyday life. Further work is required to improve the ambient noise reduction e.g. by active noise cancellation.

**Acknowledgment:** The authors would like to thank G. Ardelt for supporting this work.

**Funding:** This publication is a result of the ongoing research within the LUMEN research group, which is funded by the German Federal Ministry of Education and Research (BMBF, FKZ 13EZ1140A/B).

**Author's Statement**
Conflict of interest: Authors state no conflict of interest. Material and Methods: Informed consent: Informed consent has been obtained from all individuals included in this study. Ethical approval: The research related to human use has been complied with all the relevant national regulations, institutional policies and in accordance the tenets of the Helsinki Declaration, and has been approved by the authors' institutional review board or equivalent committee.

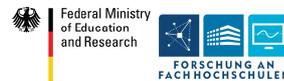